\documentclass[multphys,vecphys]{svmult}

\usepackage{makeidx}         
\usepackage{graphicx}        
\usepackage{multicol}        
\usepackage[bottom]{footmisc}

\makeindex             

\begin{document}

\title*{The source of variable optical emission is localized
  in the jet of the radio galaxy 3C\,390.3}
\titlerunning{The source of variable optical emission is localized in
  the jet of 3C\,390.3}

\author{Tigran G.~Arshakian\inst{1}, Andrei P.~Lobanov\inst{1},
  Vahram H.~Chavushyan\inst{2,3}, \\Alla I.~Shapovalova\inst{4} \and
  J.Anton~Zensus\inst{1}}
\authorrunning{T.G.~Arshakian et al.}

\institute{MPIfR, Auf dem H\"ugel 69, 53121 Bonn, Germany
\texttt{tigar@mpifr-bonn.mpg.de}
\and 	 INAOE, Apartado Postal 51 y 216, 72000 Puebla, Pue,
   M\'exico  
	     \and UNAM, Apartado Postal 70-264, 04510 M\'exico D.F.,
	     M\'exico 
	     \and SAO RAS, Nizhnij Arkhyz, Karachaevo-Cherkesia 369167,
	     Russia}

%
%
\maketitle


\section{The link between variable radio
  emission of the jet and optical continuum emission}
\label{sec:1}
The strong continuum emission is believed to originate in the central
pc-scale region of active galactic nuclei (AGN) and is responsible for
ionizing the cloud material in the broad-line region (BLR). Locating
the source of the variable continuum emission in AGN is therefore
central for understanding the mechanism for release and transport of
energy in active galaxies. In radio-loud AGN, the continuum emission
from the relativistic jets dominates at all energies. The presence of
a positive correlation between beamed radio luminosity of the jet and
optical nuclear luminosity in the sample of radio galaxies suggests
that the optical emission is non-thermal and may originate from a
relativistic jet \cite{ar:hardcastle00,ar:chiaberge02}. The detection
of a correlation between the variations of radio and optical nuclear
emission in a single source would be the most direct evidence of
optical continuum emission coming from the jet.\\

To search for a relation between variability of the optical continuum
flux and radio flux density in AGN on scales of $\sim 1$\,pc, we
combine \cite{ar:arshakian05} the results from monitoring of the
radio-loud broad emission-line galaxy 3C\,390.3 ($z=0.0561$) in the
optical
and ultraviolet 
regimes with ten very long baseline interferometry (VLBI) observations
of its radio emission at 15\,GHz \cite{ar:kellermann04} made from 1992
to 2002 using the VLBA (Very Long Baseline Array).

\begin{figure}[t]
\centering
\includegraphics[height=4.3cm,width=4.8cm]{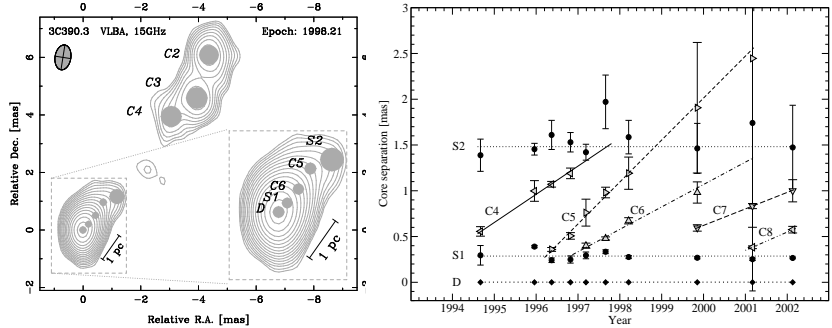}
\includegraphics[height=4.3cm,width=6cm]{arshakianF1b.eps}
%
%
\caption{{\it Left}: A single epoch (1998.21) radio structure of
  3C\,390.3 observed with VLBI at 15 GHz (1\,mas = 1.09\,pc). The labels
  in the inset mark three stationary features (D, S1 and S2) and two
  moving components (C5 and C6) identified in the jet. Images of later
  epochs showed also new components C7 and C8. {\it Right}:
  Separation of the jet components relative to the stationary feature
  D (filled diamonds) for ten epochs of VLBI observations. Moving
  components are denoted by triangles (C4-C8),
  and stationary components S1 and S2 are marked by filled circles.}
\label{ar:fig1}       
\end{figure}

The modelfitting of a single epoch VLBA image (Fig.~\ref{ar:fig1},
left panel) shows five radio components on the scale of 2 mas.  For
ten VLBA images, we identified five moving components (C4-C8) and two
stationary components (S1 and S2) separated from D by
($0.28\pm0.03$)\,mas and ($1.50\pm0.12$)\,mas, respectively
(Fig.~\ref{ar:fig1}, right panel). The proper motions of moving
components correspond to apparent speeds of $0.8\,c$ to $1.5\,c$. We
also measured the epoch, $t_{\rm D}$, at which each component was
ejected from the component D and the epoch, $t_{\rm S1}$, when it
passed through the location of the stationary feature S1
(Fig.~\ref{ar:fig1}, right panel).

%
%
\begin{figure}
\centering
\includegraphics[height=4.8cm,width=0.56\textwidth]{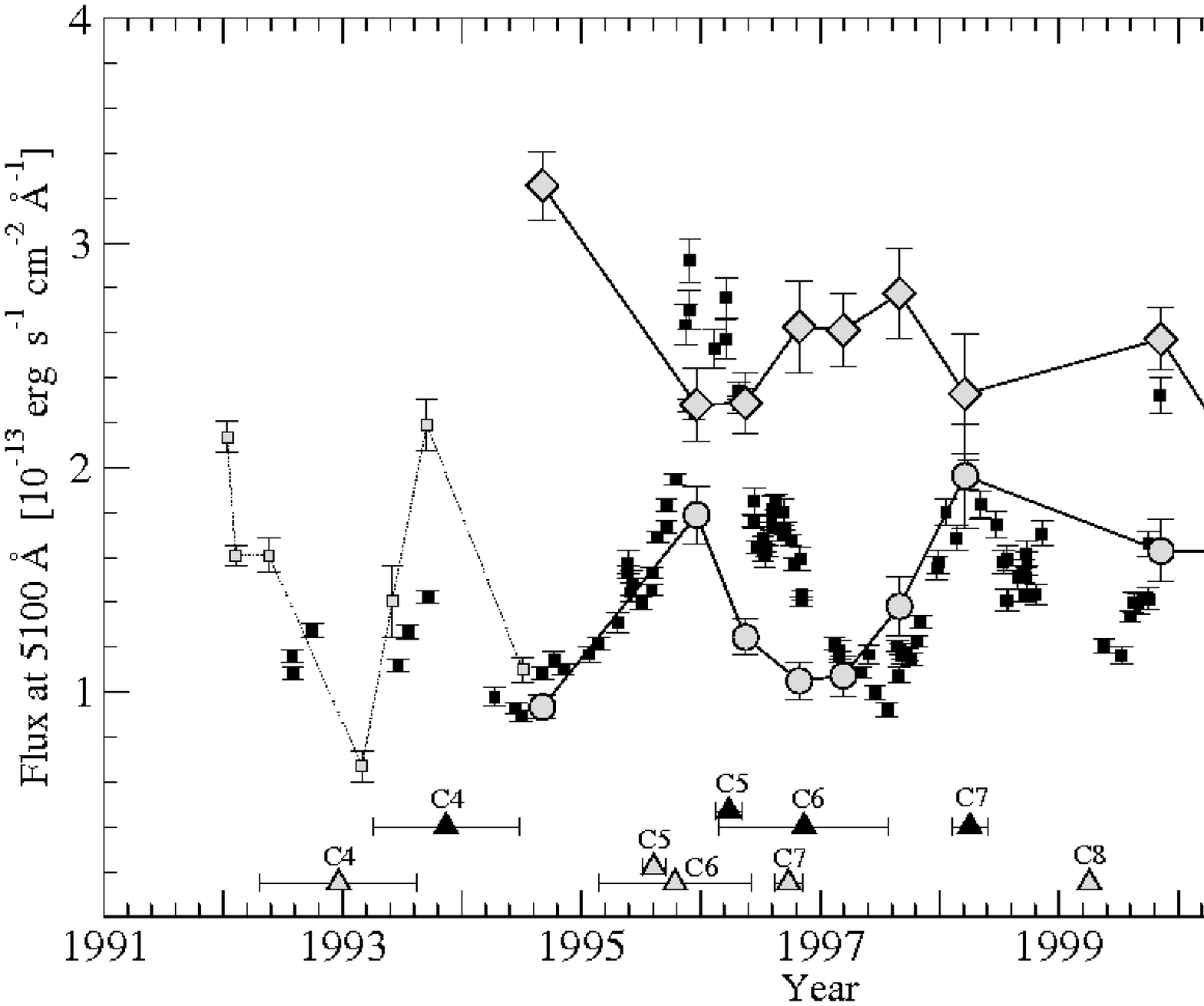}
\includegraphics[height=4cm,width=0.42\textwidth]{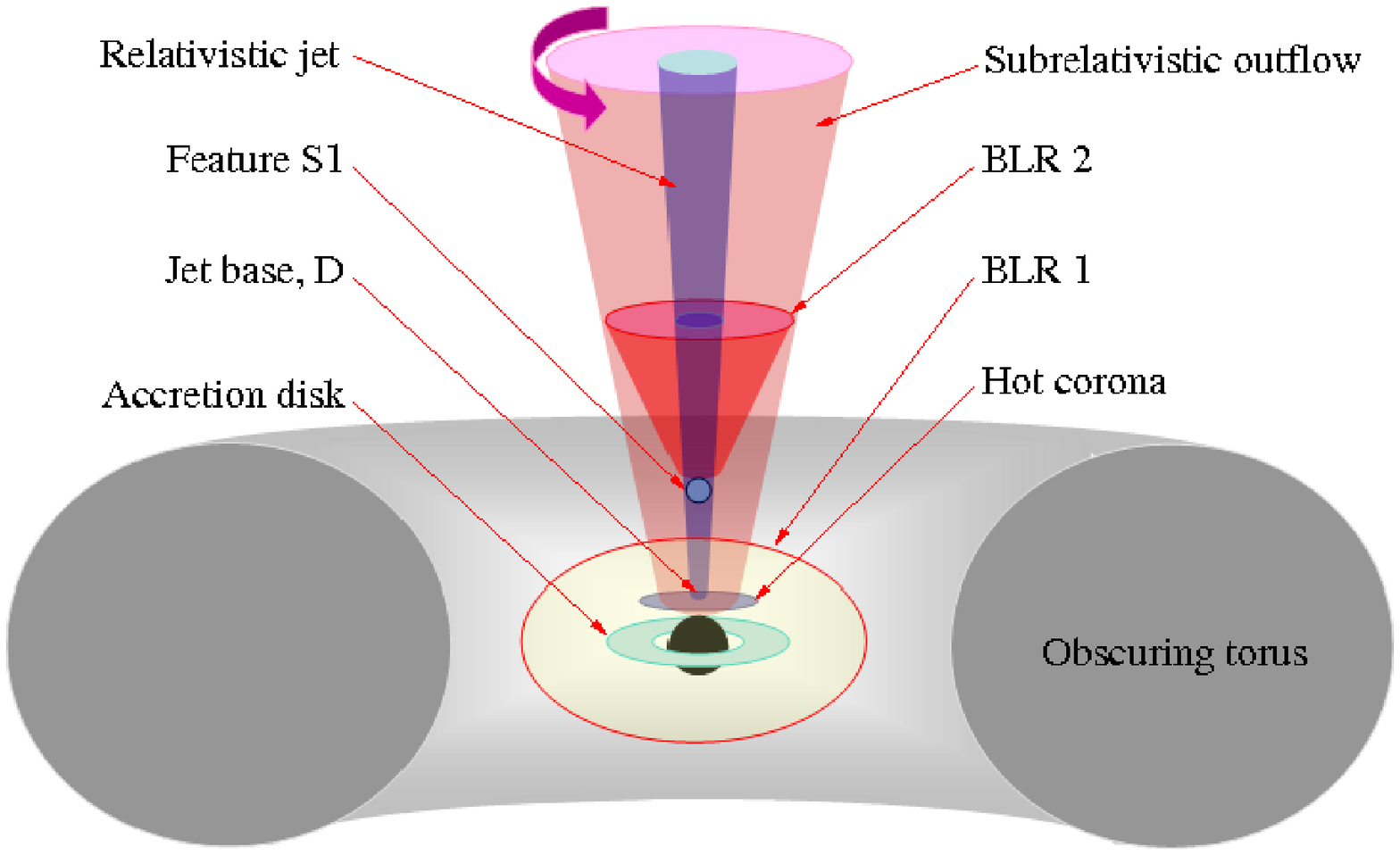}
%
%
\caption{{\it Left}: Optical continuum fluxes at 5100\AA\ (black
  squares) and 1850\AA\ (open squares) superimposed with the flux
  density at 15 GHz from the stationary components D (gray diamonds)
  and S1 (gray circles) of the jet and the epochs $t_{\rm D}$ and
  $t_{\rm S1}$ marked by open and filled triangles
  respectively. {\it Right}: A sketch of the nuclear region in 3C
  390.3 which shows only the approaching jet. The broad-line emission
  is likely to be generated both near the central nucleus (BLR\,1,
  ionized by the emission from a hot corona or the accretion disk) and
  near the S1 in a rotating subrelativistic outflow
  \cite{ar:proga00} surrounding the jet (BLR\,2, ionized by the
  emission from the relativistic plasma in the jet).}
\label{ar:fig2}       
\end{figure}

Employing minimization methods we found significant correlations
between variations in the radio emission from D and S1 and in the
optical continuum emission (Fig.~\ref{ar:fig2}, left panel). A time
delay of $\sim 1$ year was found between flux density variations of D
and S1 components of the jet. We also showed that the radio light curves
of D and S1 correlate significantly ($\ge 95$ \% confidence level) and
lead the optical light curve with time delays of $\sim 1.4$ yr and $<
0.4$ yr respectively. This finding indicates that \emph{there is a
physical link between the jet components and optical continuum: the
variable optical continuum emission is located in the innermost part of the
sub-ps-scale jet near to component S1}. A large fraction of the
variable optical continuum emission in 3C 390.3 is non-thermal
\cite{ar:wamsteker} and produced in a region of 0.004 pc in size
\cite{ar:dietrich} at a distance of $> 0.4$ pc from D
\cite{ar:arshakian05}.

The link between the optical continuum and the jet is also supported
by the correlation between the local maxima in the optical continuum
light curve and the epochs $t_{\rm S1}$ (Fig.~\ref{ar:fig2}, left
panel). The null hypothesis that this happens by chance is rejected at
a confidence level of 99.98\%. This suggests that \emph{radio
component ejection events are coupled with the long-term variability
of optical continuum}.

\section{The central sub-pc-scale region in 3C\,390.3}
\label{sec:2}
To understand the structure of the central engine and its radiation
mechanism one needs to locate the stationary features with respect to
the central nucleus. The component D is identified with the base of
the jet which is most likely to be located in the central engine of
3C\,390.3 near the accretion disk or hot corona. This is supported by
the link between the ejection epoch ($t_{\rm D}$) of the component C5
and the dip in the X-ray emission (similar to 3C\,120;
\cite{ar:marscher02}). The feature S1 can be associated with the
stationary radio feature which may be produced by internal oblique
shock formed in the continuous relativistic flow (Fig.~\ref{ar:fig2},
right panel). We suggest that the beamed synchrotron radio emission
from S1 produces the optical continuum emission via the Inverse
Compton effect. The variable optical continuum emission associated
with the jet and counterjet forms two conical shaped BLRs (BLR2;
Fig.~\ref{ar:fig2}, right panel) at a distance of about 0.4 pc from
the central nucleus.

The presence of the BLR2 at a large distance from the central nucleus
challenges \cite{ar:arshakian05} the existing models of BLRs and the
assumption of virialized motion in the BLR and hence, the BH mass
estimates \cite{peterson02} of radio-loud AGN.

%





\begin{thebibliography}{99.}

\bibitem{ar:hardcastle00} M.J. Hardcastle, D.M. Worrall: MNRAS, 314,
  359 (2000)
\bibitem{ar:chiaberge02} M. Chiaberge, A. Capetti, A. Celotti: AA,
  394, 791 (2002)
\bibitem{ar:arshakian05} T.G. Arshakian, et al: astro-ph/0512393 (2005)
\bibitem{ar:kellermann04} K.I. Kellermann, et al.: ApJ, 609, 539
  (2004)
\bibitem{ar:wamsteker} W. Wamsteker, W. Ting-gui, N. Schartel, R. Vio:
  MNRAS, 288, 225 (1997)
\bibitem{ar:dietrich} M. Dietrich, et al.: ApJSS, 115, 185 (1998)
\bibitem{ar:proga00} D. Proga, J.M. Stone, T.R. Kallman: ApJ, 543, 686 (2000)
\bibitem{ar:marscher02} A.P. Marscher, et al.: Nature, 417, 625 (2002)
\bibitem{peterson02} B.M. Peterson: Variability of AGN. In
  \textit{Advanced Lectures on The Starburst-AGN Connection}, ed by
  I. Aretxaga, D. Kunth, R. M\'ujica (Singapore World Scientific 2002)
  pp 3--68

\end{thebibliography}
\end{document}